\documentclass[11pt,a4paper]{article}
\usepackage{amsmath,amssymb}
\usepackage{epsfig,graphicx}

\def\beq{\begin{eqnarray}}    
\def\eeq{\end{eqnarray}}      

\def\ln{\,\mbox{ln}\,}                  

\def\im{\textrm{i}}

\def\diff{\textrm{d}}
\def\sfrac#1#2{{\textstyle\frac{#1}{#2}}}
\def\={\ =\ }
\def\und{\qquad\textrm{and}\qquad}

\def\al{\alpha}

\def\de{\delta}

\def\vp{\varepsilon}

\begin{document}

\title{\bf Gauge dependence of vacuum expectation values
of gauge invariant operators from soft breaking of BRST symmetry.
Example of Gribov-Zwanziger action}
\author{
P.~M.~Lavrov$^{a}$\footnote{{\bf e-mail}: lavrov@tspu.edu.ru},
A.~A.~Reshetnyak$^{a,b}$\footnote{{\bf e-mail}: reshet@ispms.tsc.ru}
\\
$^a$\small{\em Tomsk State Pedagogical University,} \small{\em Tomsk 634041,
Russia}\\
$^b$\small{\em Institute of
 Strength Physics and Materials Science,} \small{\em 634021 Tomsk, Russia}
}
\date{}
\maketitle

\begin{abstract}
We review the study of  influence of the so-called soft BRST
symmetry breaking  within the Batalin-Vilkovisky (BV)
 formalism  introduced in our papers [JHEP 1110 (2011) 043,
arXiv:1108.4820 [hep-th], MPLA 27 (2012) 1250067, arXiv:1201.4720
[hep-th]] on gauge dependence of the effective action and vacuum
expectation values of gauge invariant operators.
 We derive the Ward identities for generating
 functionals  of Green's functions for a given theory with soft
 BRST symmetry breaking  term being added to
 the quantum action and investigate  theirs gauge dependence.
 It is strongly argued that gauge theories with a soft breaking
 of BRST symmetry are inconsistent within the BV
 formalism because of the gauge-dependence of $S$-matrix.
 The application to the  Gribov-Zwanziger action (enlarging
 $SU(N)$ Yang-Mills gauge theory by means
 of not gauge-invariant horizon function)
 for the one-parameter family of $R_\xi$ gauges with
 use of the new form of the Hermitian augmented
 Faddeev-Popov operator (being by Faddeev-Popov operator for transverse components of Yang--Mills fields) is considered.

\end{abstract}

\section{Introduction}\label{intro}

\noindent

The BRST symmetry concept, equivalently  expressing a gauge
invariance via a special one-parameter global supersymmetry
\cite{brst}, not only appears as a defining tool within quantum
gauge theory, because of all known fundamental interactions are
described in terms of gauge theories \cite{books}, but provides  the
success of perturbative calculations at high energy  and of
numerical studies in lattice gauge theory \cite{lattice} as well as
 strong evidence that the interactions of quarks and gluons
are correctly described by the non-Abelian gauge theory known as
QCD.

Not long ago, the Gribov-Zwanziger (GZ)theory
\cite{Gribov,Zwanziger1,Zwanziger2}  has been intensively studied in
a series of the papers \cite{Sorella1}, \cite{Sorella2},
\cite{Sorella3}. The GZ theory is characterized by breakdown of BRST
symmetry due to the  Gribov copies being nothing else that
gauge-equivalent configurations that satisfy the Landau gauge
condition. The analytical proof \cite{Gribov} of the presence of
Gribov copies in physical spectrum was confirmed by the lattice
simulations in some kinds of QCD models, like $SU(2)$ gluodynamics ,
(see e.g. \cite{lattice2} and references therein), being not
unexpected result due to finding the field configurations within the
same Landau gauge condition. The resolution of this problem can be
realized by an addition to the quantum action constructed by
Faddeev-Popov receipt of the special horizon functional
\cite{Zwanziger1,Zwanziger2}, which is not however BRST invariant
one.

Note, that practically all the research \cite{lattice},
\cite{Sorella1}, \cite{Sorella2},  \cite{Zwanziger1},
\cite{Zwanziger2}, \cite{lattice2} of the Gribov horizon in the YM
theories have been performed in the Landau gauge only, with except
for the some kind  of covariant gauges in \cite{Sorella3} and
Coulomb gauge \cite{Zwanziger3} in the space-time with dimensions $d
= 2, 3, 4$. In spite of that fact, there is a significant
arbitrariness in the choice of the admissible gauges, which, in
part, related to the choice of the reference frame, see  e.g.
\cite{DeWitt}. It is well known the  Green's functions depend on the
choice of the gauge  however this dependence is highly structured so
that it should be canceled for the physical quantities like
$S$-matrix. Modern considerations of the gauge independence of the
$S$-matrix for the YM theories  essentially use the BRST symmetry
\cite{LT}. Therefore, any violation of the BRST symmetry principle,
as well as any ideas of the elaboration the gauge theories  with
breakdown of this symmetry leads to serious problems with
consistency of the final field-theoretical model.

In the  work, we, first,  present the results  of gauge dependence
investigations for the general gauge theories, being more general
(like the supergravity, superstring models with open algebras and
higher-spin fields as reducible gauge theories, see e.g.
\cite{hspin1}, \cite{hspin2}, \cite{hspin3})  than the YM type
theories with the so-called soft  breaking of BRST symmetry, which
are based on the research made in  \cite{llr}, \cite{lrr} in the BV
method \cite{BV1}, \cite{BV2}. Second, we suggest the horizon
functional for the family of $R_\xi$ gauge, to be constructed for
non-Hermitian Faddeev-Popov operator having for vanishing gauge
parameter $\xi$ the same form as in case of Landau gauge
\cite{Zwanziger1}, \cite{Zwanziger2}.

The paper is organized as follows. In Section~\ref{sbBRST}, we
suggest the definition of the soft breaking of BRST symmetry  in the
BV formalism. In Section~\ref{WI} we study impossibility   to
introduce BRST-like transformations  and  derive the Ward identities
for the generating functionals of Green's functions. Investigation
of the dependence of the effective action on gauges is presented in
Section~\ref{GD}. A generalization of the
 GZ action for the one-parameter family of $R_\xi$
gauges is considered in Section~\ref{GZRxi}. We summarize the
results and throw light on perspectives   in
Section~\ref{conclusion}.

We use the condensed notations  of DeWitt \cite{DeWitt} and Refs.
\cite{llr}, \cite{lrr}. Derivatives with respect to sources and
antifields are taken from the left, while those with respect to
fields are taken from the right. Left (right) derivatives with
respect to fields (antifields) are labeled by a subscript~$l$~($r$).
The Grassmann parity of any homogeneous quantity $A$  is denoted as
$\varepsilon (A)$.
\\

\section{ Soft breaking opf BRST symmetry in the BV formalism}\label{sbBRST}

\noindent
 Let us consider a theory of  gauge fields $A^i$, $i=1,2,\ldots,n$,
 ($\varepsilon(A^i)=\varepsilon_i$), with an initial
action $\mathcal{S}_0=\mathcal{S}_0(A)$ to be invariant  under the gauge transformations $\delta
A^i= R^i_{\alpha}(A)\xi^{\alpha}$ with  arbitrary functions  of the space-time coordinates $\xi^{\alpha}$
($\varepsilon(\xi^{\alpha}) =\varepsilon_{\alpha}$),
 that means the presence of the Noether's
 identities
\begin{eqnarray}
\label{GIClassA}   \mathcal{S}_{0,i}(A) R^i_{\alpha}(A)=0
\qquad\textrm{for}\quad \alpha=1,2,\ldots,m\ ,\quad 0<m<n\ .
\end{eqnarray}
among the classical equations of motion,  $\mathcal{S}_{0,i} = 0$. Here, the
functions $R^i_{\alpha}(A)$ ($\vp(R^i_\al)=\vp_i{+}\vp_\al$) are
the generators of the gauge transformations and we have used  the DeWitt's
 notation $\mathcal{S}_{0,i}\equiv\delta \mathcal{S}_0/\delta A^i$.
The structure of configuration space $\{\Phi^A\}$ in the  BV
formalism depends on the type of given classical gauge theory (for
details, on reducible or (and) gauge theories with open algebras see \cite{BV1}, \cite{BV2}). Explicit contents of $\{\Phi^A\}$   is not
critical for our aims. We need in the fact
of existence of the configuration space $\mathcal{M}$ parameterized by the
fields $\Phi\ \equiv\ \{\Phi^A\}\=\{A^i, C^\alpha,  \overline{C}{}^\alpha, B^\alpha, \ldots\}$ with
$\varepsilon(\Phi^A)=\varepsilon_A$ , where the dots indicate the
full set of additional to the classical $A^i$, ghost $C^\alpha$,   antighost $\overline{C}{}^\alpha$, Nakanishi-Lautrup $ B^\alpha$ fields in the BV method. The BV method implies an introduction to
each field $\Phi^A$ of the total configuration space the respective antifield~$\Phi^*_A$ with  opposite
Grassmann parities to that of the corresponding field $\Phi^A$,
$\Phi^*\ \equiv\ \{\Phi^*_A\} \= \{A^*_i, C^*_\alpha,  \overline{C}{}^*_\alpha, B^*_\alpha, \ldots\}$, with
$\vp(\Phi^*_A)=\vp_A{+}1$.

On the field-antifield  space  of  $(\Phi^A, \Phi^*_A)$,
(in particular, being by odd cotangent bundle $\Pi T^* \mathcal{M}$) one defines the main object of the BV quantization, i.e. the  bosonic functional ${\bar S}={\bar S}(\Phi,\Phi^*)$
obeying the master equation
\beq \label{MastEBV}
\sfrac {1}{2} ({\bar S},{\bar
S})\=\im\hbar\,{\Delta}{\bar S} \eeq
with the boundary condition
\beq \label{BoundCon} {\bar S}|_{\mathcal{M}, \hbar = 0}\ =
\mathcal{S}_0(A)\ . \eeq
to be compatible with the Eq. (\ref{MastEBV}).%

pIn (\ref{MastEBV}) we used the notation of odd second order  nilpotent Laplacian operator $\Delta$,
\beq \label{DefAB}
\Delta\ \equiv\ (-1)^{\vp_A}
\frac{\delta_{\it l}} {\delta\Phi^A}\;\frac{\delta}
{\delta\Phi^*_A},\qquad \vp
(\Delta)=1,  \eeq
and the antibracket, $(H,G)$, which may be reproduced by  $\Delta$ acting
on the product of two functionals $H$ and $G$ on field-antifield space:
\beq
\Delta\;(H \cdot G)=(\Delta H)\cdot G+ (-1)^{\varepsilon(H)}H\cdot \Delta G+
(H,G)^{\varepsilon(H)}.
\eeq

The action ${\bar S}$ should be modified by means of  corresponding fermionic gauge
fixing functional $\Psi = \Psi(\Phi)$ in such a way that we able to construct the
non-degenerate (on space $\mathcal{M}$) action $S_{ext}$ by the rule
\beq \label{ExtActBV}
S_{ext}(\Phi, \Phi^*) \= {\bar S}\big(
\Phi,\,\Phi^* + \sfrac{\de\Psi}{\de\Phi} \big)\ . \eeq
The quantum action $S_{ext}$ obeys the same master equation
(\ref{MastEBV}) as the functional ${\bar S}$,
\beq
\label{ClMastEBVExt} \sfrac {1}{2} (S_{ext}, S_{ext})\=
\im\hbar\,{\Delta}{S_{ext}}
\eeq
and should be used to construct the
generating functional of Green's functions in the BV
formalism~\cite{BV1}, \cite{BV2}.

Following to Refs. \cite{Zwanziger1}, \cite{Zwanziger2} and  our research \cite{llr}, \cite{lrr}, we deform
the action $S_{ext}$ by adding a functional $M=M(\Phi,\Phi^*)$,
defining now the full action $S$ as
\beq \label{Sfull} S\=S_{ext}+M\, \quad \vp(M)=0 .
\eeq
We speak on a \emph{soft breaking of BRST symmetry} in the
BV formalism if the condition holds%
\beq \label{SoftBrC}
\sfrac{1}{2}(M,M)\=-\im\hbar\,{\Delta}{M}. \eeq
 Note,
that in classical limit,  $\hbar\rightarrow 0$, we assume that
$M=M_0+O(\hbar)$, Eq. (\ref{SoftBrC}) is reduced to
\beq
\label{ClMasEqM} (M_0,M_0)=0. \eeq
used, in fact in Ref. \cite{llr}.
The reason to use the notation
of "a soft breaking of BRST symmetry`` in BV formalism may be explained as
follows. The master equation (\ref{ClMastEBVExt}) in the BV
formalism may  be equivalently presented in the form
\beq
\Delta \exp \Big\{\frac{\im}{\hbar}S_{ext}\Big\}=0.
\label{MEexp}
\eeq
Using the action $S_{ext}$ as a solution to this equation in order
to construct Green's functions for general gauge theories one can
derive the BRST symmetry transformations \cite{BV1}, \cite{BV2}.
Modifying the action $S_{ext}$ by a {\it special} functional $M$
(it allows us to speak of ''soft'') which satisfies the equation
\beq \label{SBRSTexp}\Delta \exp
\Big\{-\frac{\im}{\hbar}M\Big\}=0, \eeq
and is not a BRST invariant, i.e. $(S_{ext},M) - \imath \hbar M\ne 0$, we get the action $S$ from Eq. (\ref{Sfull})
not obeying the equation likes Eq.(\ref{MEexp})
\beq
\Delta \exp \Big\{\frac{\im}{\hbar}S\Big\}\neq 0.
\eeq
The BRST symmetry will be broken if we shall construct Green's
functions in the BV formalism using this action (see
beginning of the Section~\ref{WI} for details). From (\ref{ClMastEBVExt}) and
(\ref{SoftBrC}) it follows that the basic equation of our approach
to the soft breaking of BRST symmetry reads %
\beq \label{CBasEq}
\sfrac{1}{2}(S,S)-\im\hbar\,{\Delta}{S}\=(S,M)\ .
\eeq
In classical limit, for $S=S_0+\sum_{n\ge 1}\hbar^n S_n$, it follows from
(\ref{CBasEq}) the equation,
\beq \label{CBasEq0}
\sfrac{1}{2}(S_0,S_0)\=(S_0,M_0),
\eeq
coinciding in classical limit,  $\hbar\rightarrow 0$, with the basic
equation to the soft breaking of BRST
symmetry considered in Ref.\cite{llr} when  a regularization likes
dimensional one  for the local functional $S$ implies that  $\Delta
S \sim \delta(0) = 0$.

It should be  noted that the condition (\ref{SoftBrC}) will be
automatically satisfied in case when  the soft breaking of BRST symmetry
originates from a modification of the integration measure  in the
path integral. In this case  $M$ will be a functional of the field
variables $\Phi^A$ only, i.e.~$M=M(\Phi)$. As we have already
mentioned in Ref. \cite{llr}, this is exactly the situation for
Yang-Mills theory in the Landau gauge, when one takes into account the
Gribov horizon~\cite{Zwanziger1}, \cite{Zwanziger2}, \cite{Sorella2}.  We consider the more
general situation of $M=M(\Phi,\Phi^*)$  not restricting ourselves
to this special case.

Doing so, we suppose, first, that Gribov horizon may exist for general gauge theories. Second,
Gribov region of the fields $A^i$ can be singled out by an addition of the functional  $M$ to
full action of a given gauge system, but it  violates  BRST symmetry.

It is interesting to show that the right-hand side of the basic
 equation~(\ref{CBasEq}) can be presented in the form%
 \beq\label{brstop}
 (S,M)\={\hat s}M -\im\hbar\,{\Delta}M\ ,\texttt{ for }{\hat
s}\=(S_{ext},\bullet)-\im\hbar\,{\Delta}\eeq%
being the quantum Slavnov-Taylor operator defined as
$\hbar$-deformation of its classical analog.
Because of the master equation for $S_{ext}$ (\ref{ClMastEBVExt}) this operator is
nilpotent,
\beq
{\hat s}^2\=0\ .
\eeq
However, as compared to the
consideration in \cite{llr}, the presence of the additional term to
${\hat s}$ in the right-hand side of the relation
(\ref{brstop}) leads to the inequality \beq {\hat
s}\,\Big\{\sfrac{1}{2}(S,S)-\im\hbar \Delta S\Big\}\ \ne 0\ . \eeq
which being written for the action $S_{ext}$ should be the identity
for general gauge theories {\it without\/} a (soft) breaking of BRST symmetry.

\section{Modifications of BRST transformations and Ward identities}\label{WI}

\noindent Here we  consider some quantum consequences of the
 equations (\ref{ClMastEBVExt}), (\ref{SoftBrC})
and~(\ref{CBasEq}). To do it we introduce the generating
functional of Green's functions,
\beq \label{ZBV} Z(J,\Phi^*)\=\int\!D\Phi\ \exp
\Big\{\frac{\im}{\hbar} \big(S(\Phi,\Phi^*)+ J_A\Phi^A\big)\Big\}\ ,
\eeq
with $J_A$ ($\vp(J_A) = \vp_A$)
being  the usual sources for the fields~$\Phi^A$ and $S(\Phi,\Phi^*)$ being by a solution of the basic quantum equation
(\ref{CBasEq}) and having the form~(\ref{Sfull}).

The integrand of the vacuum functional $Z(0,\Phi^*) = Z(\Phi^*)$
 looks as
\beq \label{intgM}
{\cal N}={\cal N}(\Phi,\Phi^*)=D\Phi\ \exp
\Big\{\frac{\im}{\hbar} S(\Phi,\Phi^*)\Big\}. \eeq
Remind that in the BV formalism the BRST symmetry appears as invariance of the
integrand ${\cal N}$ of vacuum functional under the change of variables,
$\Phi^A \to \Phi^{\prime A} = \Phi^A +\delta_B \Phi^A  $
determined with the help of the non-BRST broken action $S_{ext}(\Phi,\Phi^*)$,
\beq \label{BRST}
\delta_B \Phi^A = \frac{\delta  S_{ext}}{\delta
\Phi^*_A}\theta, \qquad  \delta_B \Phi^*_A = 0,
\eeq
where $\theta$ is a nilpotent constant odd parameter. Carrying out
the change of variables (\ref{BRST}) in (\ref{intgM}) we obtain
\beq \label{Zvacnbrst} {\cal N}^{'}= {\cal N}\Big(1 -
\frac{\im}{\hbar}\theta \;\frac{\delta M}{\delta \Phi^A}\frac{\delta
S_{ext}}{\delta \Phi^*_A}  \Big). \eeq
Non-invariance of the integrand means violation of the standard
BRST symmetry. Of course, one may to restore the invariance of $\mathcal{N}$ by  modifying of the definition of BRST transformations in case of the theory under consideration. A
general
 way  for its modification
may be achieved with help of following one-parameter functional $S_{\kappa}$
\beq
S_{\kappa}=S_{ext}+\kappa M, \quad \kappa \in \mathbb{R},
\eeq
to define the BRST trasformations
\beq
\label{nmodBRST}
\delta_{B_{\kappa}} \Phi^A = \frac{\delta  S_{\kappa}}{\delta
\Phi^*_A}\theta, \qquad  \delta_{B_{\kappa}} \Phi^*_A = 0.
\eeq
Note, we have  the standard BRST
transformations (\ref{BRST}) for $\kappa=0$. Performing the change of variables (\ref{modBRST})  in the
integrand  ${\cal N}$ (\ref{intgM}) one has
\beq \label{ZBVnmodbrst} {\cal N}^{'}= {\cal N}\Big(1 -
\frac{\im}{\hbar}\theta \; \Big[\kappa (S_{ext}, M)\ -2
\im\hbar\,\kappa {\Delta}M + (1-\kappa)\frac{\delta M}{\delta
\Phi^A}\frac{\delta S_{ext}}{\delta \Phi^*_A} \Big]\Big ) . \eeq
So that we may state, that the non-invariance of the integrand exists
for any choice of the parameter $\kappa$. For instance, for $\kappa =1$
we have the modified BRST-like transformations
\beq
\label{modBRST}
\delta_B \Phi^A = \frac{\delta  S}{\delta
\Phi^*_A}\theta, \qquad  \delta_B \Phi^*_A = 0,
\eeq
which is not stay invariant the integrand $\mathcal{N}$.

Let us turn to the properties of the generating functional
$Z(J,\Phi^*)$. First, from the averaging of the Eq. (\ref{MEexp})  over the total configuration space $\mathcal{M}$  with measure $\exp
\big\{\frac{\im}{\hbar} \big(S+J_A\Phi^A\big)\big\}$
\beq \nonumber 0 \= \int\!D\Phi\ \Delta \exp
\Big\{\frac{\im}{\hbar}S_{ext}\Big\}\; \exp \Big\{\frac{\im}{\hbar}
\big(S+J_A\Phi^A\big)\Big\}\ \eeq
it follows  after integrating by parts in the functional integral the
following identity for the generating functional $Z$,
\beq
\label{WIZBV}
&&\Big(J_A+M_{A}\big(\sfrac{\hbar}{\im}\sfrac{\delta}{\delta
J},\Phi^*\big)\Big)\left(\frac{\hbar}{\im}\frac{\delta
}{\delta\Phi^*_A}\ -\
M^{A*}\big(\sfrac{\hbar}{\im}\sfrac{\delta}{\delta
J},\Phi^*\big)\right)Z(J,\Phi^*)=0.
\eeq
In the Eq. (\ref{WIZBV})  the notations
\beq \nonumber
M_{A}\big(\sfrac{\hbar}{\im}\sfrac{\delta}{\delta
J},\Phi^*\big)\equiv \frac{\delta M(\Phi,\Phi^*)}{\delta
\Phi^A}\Big|_{\Phi\rightarrow
\frac{\hbar}{\im}\frac{\delta}{\delta J}} \und
M^{A*}\big(\sfrac{\hbar}{\im}\sfrac{\delta}{\delta
J},\Phi^*\big)\equiv \frac{\delta
M(\Phi,\Phi^*)}{\delta\Phi^*_A}\Big|_{\Phi\rightarrow
\frac{\hbar}{\im}\frac{\delta}{\delta J}}
\eeq
have been used. In case of $M=0$, the identity  (\ref{WIZBV}) is
reduced to the usual Ward identity for the generating functional of
Green's functions in the BV formalism. Therefore, we refer
to~(\ref{WIZBV}) as the Ward identity for~$Z$ in a gauge theory with
softly broken BRST symmetry. Note, that for the regularization
scheme likes dimensional one, we will have $\Delta M = 0$ and,
therefore the equation (\ref{SoftBrC}) is reduced to
$\big(M,\,M\big)=0$, which leads to the vanishing of the
combination, $M_{A}\big(\sfrac{\hbar}{\im}\sfrac{\delta}{\delta
J},\Phi^*\big) M^{A*}\big(\sfrac{\hbar}{\im}\sfrac{\delta}{\delta
J},\Phi^*\big)$ in the Ward identity (\ref{WIZBV}) as it was firstly derived
in \cite{llr}.

Second, after introducing the generating functional of connected Green's
functions, %
\beq W(J,\Phi^*) \= -\im\hbar \ln Z(J,\Phi^*)\ ,%
\eeq
the Ward identity (\ref{WIZBV}) can be rewritten for $W$ as %
\beq \label{WIWBV}
 &&\Big(J_A+M_{A}\big(\sfrac{\delta W}{\delta J}+
\sfrac{\hbar}{\im}\sfrac{\delta}{\delta
J},\Phi^*\big)\Big)\left(\frac{\delta
W(J,\Phi^*)}{\delta\Phi^*_A}\ -\ M^{A*}\big(\sfrac{\delta
W}{\delta J}+\sfrac{\hbar}{\im}\sfrac{\delta}{\delta
J},\Phi^*\big)\right)=0 . \eeq

In turn, the generating functional of the vertex functions (effective
action) is obtained by Legendre transforming of $W$, \beq
\label{EA} \Gamma (\Phi,\,\Phi^*) \= W(J,\Phi^*) - J_A\Phi^A,
\qquad\textrm{where}\quad \Phi^A = \frac{\delta W}{\delta J_A},
\qquad \frac{\delta\Gamma}{\delta\Phi^A}=-J_A\ . \eeq
Taking into
account the equality, $\frac{\delta \Gamma}{\delta
\Phi^*_A}\=\frac{\delta W}{\delta\Phi^*_A}$, which follows from the Legendre transformation,
 we can present the identity
(\ref{WIWBV}) in terms of~$\Gamma$ as
\beq \label{WIGammaBV}
\sfrac{1}{2}(\Gamma,\Gamma) \=
\frac{\delta\Gamma}{\delta\Phi^A}{\widehat M}^{A*}
+{\widehat M}_{A}\frac{\delta \Gamma}{\delta
\Phi^*_A}-{\widehat M}_{A}{\widehat M}^{A*}\ .
\eeq
In the Eq. (\ref{WIGammaBV}) we have used the notations \beq {\widehat M}_{A}\ \equiv\
\frac{\delta
M(\Phi,\Phi^*)}{\delta\Phi^A}\Big|_{\Phi\to\widehat\Phi} \und
{\widehat M}^{A*}\ \equiv\ \frac{\delta
M(\Phi,\Phi^*)}{\delta\Phi^*_A}\Big|_{\Phi\to\widehat\Phi}\,, \eeq
where the sign ${\widehat\Phi}^A$ means the field $\Phi^A$
enlarged  by the special  derivatives $\im\hbar\,(\Gamma^{''-1})^{AB}\frac{\delta_l}{\delta\Phi^B}$, %
 \beq
{\widehat\Phi}^A\=\Phi^A+\im\hbar\,(\Gamma^{''-1})^{AB}
\frac{\delta_l}{\delta\Phi^B} \eeq and the matrix
$(\Gamma^{''-1})$ is inverse to the matrix $\Gamma^{''}$ with
elements \beq (\Gamma^{''})_{AB}\=\frac{\delta_l}{\delta\Phi^A}
\Big(\frac{\delta\Gamma}{\delta\Phi^B}\Big)\
: \quad
(\Gamma^{''-1})^{AC}(\Gamma^{''})_{CB}=\delta^A_{\ B}\ . \eeq
We
see again, in  case of vanishing $M$ the identity (\ref{WIGammaBV})
coincides with the Ward identity for the effective action in the
BV formalism. Emphasize that the identity (\ref{WIGammaBV}) is
compatible with the classical equation (\ref{CBasEq}), since
$\hbar\to0$ yields $\Gamma=S_0$, ${\widehat M}=M_0$, and
(\ref{WIGammaBV})  is reduced to~(\ref{CBasEq0}).

In similar manner  we can derive the Ward identity
which follows from (\ref{SBRSTexp}). To this end, we average the
equation (\ref{SBRSTexp})  over the  configuration space of the
fields
 $\Phi^A$ with measure $\exp
\big\{\frac{\im}{\hbar} \big(S_{ext}+J_A\Phi^A\big)\big\}$,
 \beq
\nonumber 0 \= \int\!D\Phi\ \Delta \exp
\Big\{-\frac{\im}{\hbar}M\Big\}\; \exp \Big\{\frac{\im}{\hbar}
\big(S_{ext}+J_A\Phi^A\big)\Big\}\ . \eeq
and derive after usual manipulations with functional integral the
identity in terms of mean fields $\Phi^A$ (\ref{EA}) %
 \beq \label{WIM}{\widehat M}_{A}{\widehat M}^{A*}=-i\hbar {\widehat
M}_A^{\;\;A^*}
\texttt{ where }
{\widehat M}_A^{\;\;A^*}=
\frac{\delta^2 M}{\delta\Phi^*_A\delta\Phi^A}\Big|_{\Phi\to\widehat\Phi}.
\eeq
 The identity (\ref{WIM})  is reduced to the
identity, ${\widehat M}_{A}{\widehat M}^{A*}=0$, derived in
\cite{llr}, when the regularization scheme likes dimensional one
is applied.
\\

\section{Gauge dependence of the effective action}\label{GD}

\noindent In this Section we present our research \cite{llr}, \cite{lrr} devoting to the gauge dependence of the generating
functionals $Z$, $W$ and $\Gamma$ for general gauge theories with
a soft breaking of BRST symmetry as it was defined in the previous section. The derivation of this dependence is based on the fact that
any variation of the gauge-fixing functional,
$\Psi(\Phi)\rightarrow\Psi(\Phi)+\delta\Psi(\Phi)$, leads to a
variation both the action $S_{ext}$ (\ref{ExtActBV}), the
functional $Z$~\cite{VLT} and the functional $M$. The variation of
$S_{ext}$ can be presented in the form
\beq \label{varSext} \delta
S_{ext}\=\frac{\delta \delta\Psi}{\delta\Phi^A}\,\frac{\delta
S_{ext}}{\delta\Phi^*_A}\qquad \texttt{ or as } \qquad \delta
S_{ext}\=-(S_{ext},\delta\Psi)\= -{\hat s}\,\delta\Psi\ ,
\eeq
 whereas we denote as $\delta M(\Phi,\Phi^*)$
the variation of $M$ corresponding to the variation~$\delta\Psi$.
 From (\ref{ZBV}), (\ref{varSext}) and the variation of $M$
we obtain the gauge variation of~$Z$,
\beq \label{varZ}
\delta Z(J,\Phi^*)\=\frac{\im}{\hbar}\int\!D\Phi\ \Big( \frac{\delta
\delta\Psi}{\delta\Phi^A}\frac{\delta S_{ext}}{\delta\Phi^*_A}+\delta
M\Big)\; \exp \Big\{\frac{\im}{\hbar}
\big(S(\Phi,\Phi^*)+ J_A\Phi^A\big)\Big\}\ .
\eeq
By means of the equality
\beq \label{AuxId} \nonumber
0&=&\int\!D\Phi\ \frac{\delta_l}{\delta
\Phi^A}\Big[\delta\Psi\;\frac{\delta S_{ext}}{\delta\Phi^*_A}\;\exp
\Big\{\frac{\im}{\hbar} \big(S(\Phi,\Phi^*)+
J_A\Phi^A\big)\Big\}\Big] \\ \nonumber
&=&\int\!D\Phi\ \Big[\frac{\delta\delta\Psi
}{\delta\Phi^A}\,\frac{\delta
S_{ext}}{\delta\Phi^*_A}-\frac{\im}{\hbar}\Big(J_A+\frac{\delta
S}{\delta \Phi^A}\Big)\frac{\delta
S_{ext}}{\delta\Phi^*_A}\,\delta\Psi\Big]\exp \Big\{\frac{\im}{\hbar}
\big(S(\Phi,\Phi^*)+ J_A\Phi^A\big)\Big\}\ ,
\eeq
where the equation (\ref{ClMastEBVExt}) was used,
we can rewrite the variation (\ref{varZ}) as
\beq \nonumber \delta Z(J,\Phi^*) &=&
\frac{\im}{\hbar}\Big[\Big(J_A+M_{A}
\big(\sfrac{\hbar}{\im}\sfrac{\delta}{\delta
J},\Phi^*\big)\Big)\left(\frac{\delta}{\delta\Phi^*_A}\,
 -\frac{\im}{\hbar}
M^{A*} \big(\sfrac{\hbar}{\im}\sfrac{\delta}{\delta
J},\Phi^*\big)\right)\,\delta\Psi
\big(\sfrac{\hbar}{\im}\sfrac{\delta}{\delta J}\big) \\
&& + \delta M\big(\sfrac{\hbar}{\im}\sfrac{\delta}{\delta
J},\Phi^*\big)\Big]Z(J,\Phi^*)\label{varZ1}.
\eeq

Now, it is easy to get the corresponding variation of the generating
functional of connected Green's functions, $\delta W(J,\Phi^*)\=
\sfrac{\hbar}{\im}Z^{-1}\delta Z$, in the form
\beq
\label{varW} \delta W(J,\Phi^*)=
\Big(J_A+M_{A}\big(\sfrac{\delta W}{\delta
J}+\sfrac{\hbar}{\im}\sfrac{\delta}{\delta
J},\Phi^*\big)\Big)\frac{\delta }{\delta\Phi^*_A} \delta\Psi
\big(\sfrac{\delta W}{\delta
J}+\sfrac{\hbar}{\im}\sfrac{\delta}{\delta J}\big) \ +\ \delta
M\big(\sfrac{\delta W}{\delta
J}+\sfrac{\hbar}{\im}\sfrac{\delta}{\delta J},\Phi^*\big) \,,
\eeq
with use of the Ward identity (\ref{WIWBV}).

Now, we are able to reach  our final purpose concerning  the
derivation of the gauge variation of the effective action. Doing so, we repeat our derivation from Ref. \cite{lrr}.

  First,
we note that $\delta \Gamma=\delta W$. Second, we observe that the
change of the variables $(J_A, \Phi^*_A) \to (\Phi^A, \Phi^*_A)$
from the Legendre transformation  (\ref{EA}) implies that
\beq
\label{dphistar}
{\frac{\delta}{\delta\Phi^*}}\Big|_{J}=
\frac{\delta}{\delta\Phi^*}\Big|_{\Phi} + \frac{\delta
\Phi}{\delta\Phi^*}{\frac{\delta_{\it
l}}{\delta\Phi}}\Big|_{\Phi^*}.
\eeq

Third, differentiating  the Ward identities for
$Z$~(\ref{WIZBV}) with respect to the sources $J_B$,  we get
\beq \nonumber
 &&\frac{\hbar}{i}\frac{\delta Z}{\delta
\Phi^*_B}+\frac{\hbar}{i}
\Big(J_A+M_{A}\big(\sfrac{\hbar}{\im}\sfrac{\delta}{\delta
J},\Phi^*\big) \frac{\delta^2 Z}{\delta
J_B\delta\Phi^*_A}(-1)^{\varepsilon_A\varepsilon_B}
 -M^{B*}\big(\sfrac{\hbar}{\im}\sfrac{\delta}{\delta
 J},\Phi^*\big)Z-\\
&&-(-1)^{\varepsilon_B}J_AM^{A*}
\big(\sfrac{\hbar}{\im}\sfrac{\delta}{\delta
J},\Phi^*\big)\frac{\delta Z}{\delta J_B}-
(-1)^{\varepsilon_B}M_{A}\big(\sfrac{\hbar}{\im}\sfrac{\delta}{\delta
J},\Phi^*\big)M^{A*} \big(\sfrac{\hbar}{\im}\sfrac{\delta}{\delta
J},\Phi^*\big)\frac{\delta Z}{\delta J_B}=0. \label{dWIZ}
\eeq
Next, using the interrelation of the  derivatives for $Z$ and $W$,
\beq && \left(\frac{\delta Z}{\delta J_A}, \frac{\delta Z}{\delta
\Phi^*_A}\right) = \frac{\im}{\hbar}\exp\{\sfrac{\im}{\hbar} W\}
\left(\frac{\delta W}{\delta J_A},  \frac{\delta W}{\delta
\Phi^*_A}\right),\\
&&   \frac{\delta^2 Z}{\delta \Phi^*_B\delta J_A}
=\exp\{\sfrac{\im}{\hbar} W\}\left[
\left(\frac{\im}{\hbar}\right)^2 \frac{\delta W}{\delta
\Phi^*_B}\frac{\delta W}{\delta
J_A}+\frac{\im}{\hbar}\frac{\delta^2 W}{\delta \Phi^*_B\delta
J_A}\right],\eeq
Eqs. (\ref{dWIZ}) may be presented in terms of functional $W$ as,
\beq \label{dW} &&  \frac{\delta
W(J,\Phi^*)}{\delta\Phi^*_B}+(-1)^{\vp_B}
\Big(J_A+M_{A}\big(\sfrac{\hbar}{\im}\sfrac{\delta}{\delta
J}+\sfrac{\delta W}{\delta J},\Phi^*\big)\Big)\frac{\delta^2
W(J,\Phi^*)}{\delta\Phi^*_A \delta J_B}\ -\
M^{B*}\big(\sfrac{\hbar}{\im}\sfrac{\delta}{\delta
J}+\sfrac{\delta W}{\delta
J},\Phi^*\big)\nonumber\\
&&  \= -\frac{\im}{\hbar}(-1)^{\vp_B}
\Big(J_A+M_{A}\big(\sfrac{\hbar}{\im}\sfrac{\delta}{\delta
J}+\sfrac{\delta W}{\delta J},\Phi^*\big)\Big) \left(\frac{\delta
W}{\delta
\Phi^*_A}-M^{A*}\big(\sfrac{\hbar}{\im}\sfrac{\delta}{\delta
J}+\sfrac{\delta W}{\delta J},\Phi^*\big)\right)\frac{\delta
W}{\delta J_B}\ . \eeq
Then, from (\ref{dW}) it follows
\beq  \frac{\delta \Gamma}{\delta\Phi^*_B}- {\widehat M}^{B^*}
-\Big(\frac{\delta\Gamma}{\delta\Phi^A}-{\widehat
M}_{A}\Big)\frac{\delta\Phi^B}{\delta\Phi^*_A}
(-1)^{\varepsilon_B}  = \frac{i}{\hbar}(-1)^{\varepsilon_B}
\Big(\frac{\delta\Gamma}{\delta\Phi^A}-{\widehat
M}_{A}\Big)\Big(\frac{\delta \Gamma}{\delta\Phi^*_A}- {\widehat
M}^{A^*}\Big)\Phi^B. \label{dG} \eeq
To make more simple the above expression one should commute the fields
$\Phi^B$ to the left in the last summand in order  to use Ward
identity for effective action $\Gamma$
(\ref{WIGammaBV})
%
As a result, we rewrite the relation (\ref{dG}) in the form
\beq \nonumber -\Big(\frac{\delta\Gamma}{\delta\Phi^A}-{\widehat
M}_{A}\Big)\frac{\delta\Phi^B}{\delta\Phi^*_A}&=&
-\Big(\frac{\delta\Gamma}{\delta\Phi^*_B}-{\widehat
M}^{B^*}\Big)(-1)^{\varepsilon_B}\\\label{dGamma}
&&+\frac{i}{\hbar}\Big[-{\widehat
M}_{A}\frac{\delta\Gamma}{\delta\Phi^*_A}-
\frac{\delta\Gamma}{\delta\Phi^A}{\widehat M}^{A^*}+{\widehat
M}_{A}{\widehat M}^{A^*},\Phi^B\Big\}, \eeq
where the  brackets $\big[\ ,\ \big\}$ denote  the
supercommutator.

From (\ref{varW}), (\ref{dphistar}) and (\ref{dGamma})
the variation of the effective action can be presented in the "local-like" form,
\beq \delta\Gamma &=& -(\Gamma,\langle\delta\Psi\rangle)\ +\
\left({\widehat M}_{A} \frac{\delta}{\delta \Phi^{*}_{A}} +\
(-1)^{\vp_A} {\widehat
M}^{A*} \frac{\delta_l}{\delta \Phi^A}\right)\langle\delta\Psi\rangle  \nonumber \\
&& -\ \frac{\im}{\hbar}\Big[{\widehat M}_A \frac
{\delta\Gamma}{\delta\Phi^*_A}+\frac{\delta\Gamma}{\delta
\Phi^A}{\widehat M}^{A*} -{\widehat M}_A{\widehat M}^{A*}  ,\
\Phi^B\Big\} \frac{\delta_{\it
l}}{\delta\Phi^B}\,\langle\delta\Psi\rangle\ +\ \langle\delta
M\rangle \label{varGamma}\ , \eeq
with local (for $M=0$) operator acting on the functional $
\langle\delta\Psi\rangle$. Here  we imply the notations
\beq \langle\delta\Psi\rangle\=\delta\Psi({\widehat\Phi})\cdot 1
\und \langle\delta M\rangle\=\delta M({\widehat \Phi},\Phi^*)\cdot
1 \ .\eeq
Then, using  the  identities,
\beq \frac{\delta \Phi^B}{\delta\Phi^*_A} =
(-1)^{\vp_B(\vp_A+1)}\frac{\delta }{\delta J_B} \frac{\delta W}{
\delta \Phi^*_A} =- (-1)^{\vp_B(\vp_A+1)}(\Gamma^{''-1})^{BC}
\frac{\delta_{\it l} }{\delta \Phi^C}\frac{\delta \Gamma}{ \delta
\Phi^*_A}, \eeq
following from the Legendre transformation (\ref{EA}) we can
present the variation of the effective action in the equivalent,
so-called non-local (due to explicit presence of the quantities
$(\Gamma^{''-1})^{BC}$) form,
 \beq \delta\Gamma \= \frac{\delta\Gamma}{\delta\Phi^A}
{\widehat F}^A\,\langle\delta\Psi\rangle\ -\ {\widehat
M}_A{\widehat F}^A \langle\delta\Psi\rangle\ +\ \langle\delta
M\rangle\ , \label{varGammaF} \eeq
where the operator  ${\widehat F}^A$ is derived from the Eqs.
(\ref{dphistar}), (\ref{dGamma}), (\ref{varGamma}) as follows
\beq {\widehat F}^A &=&-\frac{\delta}{\delta\Phi^*_A}\ +\
(-1)^{\vp_B(\vp_A+1)} (\Gamma^{''-1})^{BC}\Big(\frac{\delta_{\it
l}}{\delta\Phi^C}\frac {\delta
\Gamma}{\delta\Phi^{*}_{A}}\Big)\frac{\delta_{\it l}
}{\delta\Phi^B}\ . \label{FAdef} \eeq

From the variation (\ref{varGammaF}) it follows that on shell the
effective action is generally gauge dependent because of
\beq \frac{\delta\Gamma}{\delta\Phi^A}=0 \qquad
\longrightarrow\qquad \delta\Gamma\neq 0\ .
\eeq
This fact does not permit to formulate  consistently  of a soft
breaking of BRST symmetry within the field-antifield formalism, if
only  two last terms in~(\ref{varGammaF}) cancel each other,
\beq \label{BasRest}
\langle\delta M\rangle\={\widehat M}_A{\widehat F}^A \langle\delta\Psi\rangle\ .
\eeq
However, this is rather a strong restriction on the BRST-breaking
functional~$M$ for the effective action to be gauge independent
on-shell. The same statement is valid for the physical S-matrix.
Really, Eq. (\ref{BasRest})~fixes the gauge variation of
$M=M(\Phi,\Phi^*)$ under a change of the gauge-fixing
functional~$\Psi$ to be
\beq \label{BEqvM} \delta M\=\frac{\delta
M}{\delta\Phi^A}\,{\widehat F}_0^A\,\delta\Psi\quad\texttt{ where
}\quad {\widehat F}_0^A\=(-1)^{\vp_B(\vp_A+1) } (S^{''-1})^{BC}
\Big(\frac{\delta_{\it l}}{\delta\Phi^C}\frac {\delta
S}{\delta\Phi^{*}_{A}}\Big) \frac{\delta_{\it l} }{\delta\Phi^B}\
. \eeq
It was shown in \cite{llr} that already in the case of Yang-Mills
theories in linear $R_{\xi}$ gauge which includes the Landau gauge
the relation (\ref{BEqvM}) does not satisfy. We are forced to claim
that a consistent quantization of general gauge theories when
restriction on the domain of integration in functional integral  is
taken as an addition to the full action of a given gauge system
violating the BRST symmetry does not exist. As a consequence, the last fact implies that the vacuum expectation values
of gauge invariant operators calculated for the theory with soft breaking of the BRST symmetry is gauge dependent.

\section{Gribov-Zwanziger action in one-parameter $R_{\xi}$-gauges}\label{GZRxi}

In this section we shall apply our above-described general consideration
of a soft BRST breaking to the important case of Yang-Mills theories,
since those had been the subject of recent investigations~\cite{Sorella1}--\cite{Sorella3}.
The initial classical action $S_0$ of Yang-Mills fields $A^a_{\mu}(x)$,
which take values in the adjoint representation of~$su(N)$ so that,
$a=1,\ldots,N^2{-}1$, has the standard form
\beq
S_0(A) \= -\sfrac14\int\!\diff^D x\ F_{\mu\nu}^{a}F^{\mu\nu{}a}
\qquad\textrm{with}\quad
F^a_{\mu\nu}\=\partial_{\mu}A^a_{\nu}-\partial_{\nu}A^a_{\mu}+
f^{abc}A^b_{\mu}A^c_{\nu}\ , \label{clYM}
\eeq
where $\mu,\nu=0,1,\ldots,D{-}1$, the Minkowski space has mostly $"+"$ signature,
$(-,+,\ldots,+)$, and $f^{abc}$ denote the (totally antisymmetric) structure
constants of the Lie algebra~$su(N)$.
The action (\ref{clYM}) is invariant under the gauge transformations
\beq \delta A^a_{\mu}\=D^{ab}_{\mu}\xi^b \qquad\textrm{with}\quad
D^{ab}_{\mu}\=\delta^{ab}\partial_{\mu}+f^{acb}A^c_{\mu}\ .\eeq
The total field configuration space of Yang-Mills theory,
\beq \{\Phi^A\}\=\{A^a_{\mu}, B^a, C^a, {\bar C}^a\}
\qquad\textrm{with}\quad
\varepsilon(C^a)=\varepsilon(\bar C^a)=1\ ,\quad
\varepsilon(A^a_\mu)=\varepsilon(B^a)=0\ , \eeq
includes the (scalar) Faddeev-Popov ghost and antighost fields
$C^a$ and ${\bar C}^a$, respectively, as well as the
Nakanishi-Lautrup auxiliary fields $B^a$.
The corresponding set of antifields is
\beq \{\Phi^*_A\} \=\{A^{*a\mu}, B^{*a}, C^{*a}, {\bar C}^{*a}\}
\qquad\textrm{with}\quad
\varepsilon(A^{*a\mu})=\varepsilon(B^{*a})=1\ ,\quad
\varepsilon(C^{*a})= \varepsilon({\bar C}^{*a})=0\ . \eeq
A solution to the classical master equation  can be presented in the
form
\beq \bar{S}(\Phi,\Phi^*) \= S_0(A)\ +\ A^{*a\mu}D^{ab}_{\mu}C^b\ +\
\sfrac{1}{2}C^{*a}f^{abc}C^bC^c\ +\ \bar{C}{}^{*a}B^a\label{SbosGZ}\ .\eeq
The gauge-fixing functional can be chosen as
\beq \Psi(\Phi)\={\bar C}^a\chi^a(A,B) \eeq
with free bosonic functions~$\chi^a$,
so that the non-degenerate action $S_{ext}$~(\ref{ExtActBV}) becomes
\beq \nonumber
S_{ext}(\Phi,\Phi^*)&=& S_0(A) + \Big(A^{*a\mu}+{\bar
C}^c\frac{\delta\chi^c}{\delta A^a_{\mu}}\Big)D^{ab}_{\mu}C^b +
\sfrac{1}{2}C^{*a}f^{abc}C^bC^c +
\big(\bar{C}{}^{*a}+\chi^a\big)B^a \\
&=&S_{FP}(\Phi)\ +\ A^{*a\mu}D^{ab}_{\mu}C^b\ +\
\sfrac{1}{2}C^{*a}f^{abc}C^bC^c\ +\ \bar{C}{}^{*a}B^a\ ,
\label{ExtYMBV}\eeq
where $S_{FP}(\Phi)$ is the Faddeev-Popov action
\beq\label{FPact}
S_{FP}(\Phi)\=S_0(A)\ +\ {\bar C}^a K^{ab}C^b\ +\ \chi^a B^a
\qquad\textrm{with}\qquad
K^{ab}=\frac{\delta\chi^a}{\delta A^c_{\mu}}D^{cb}_{\mu}\ .
\eeq
The actions (\ref{FPact}) and (\ref{ExtYMBV}) are invariant
under the BRST transformation
\beq\label{BRSTtr}
\delta_B A_{\mu}^{a} = D^{ab}_{\mu}C^b\theta\ ,\quad
\delta_B \bar{C}{}^a = B^a\theta\ ,\quad
\delta_B B^a = 0\ ,\quad
\delta_B C^a = \sfrac12 f^{abc}C^bC^c\theta
\label{BRSTGZred} \eeq
where $\theta$ is a constant Grassmann parameter.

In \cite{Zwanziger1,Zwanziger2} it has been shown that the
Gribov horizon \cite{Gribov} in Yang-Mills theory (\ref{clYM})
in the Landau gauge,
\beq \label{Landau}
\chi^a(A,B)\=\partial^{\mu}A_{\mu}^a
\qquad\longrightarrow\qquad
K^{ab}=\partial^\mu D_\mu^{ab}\ ,
\eeq
can be taken in to account by adding to the Faddeev-Popov action (\ref{FPact})
the non-local functional~\footnote{
    The choice of \cite{Sorella1,Sorella2,Sorella3} agrees with ours after Wick rotation,
    integrating out auxiliary fields and renaming $\gamma^4\to\gamma^2$.}
\beq \label{FuncM}
M(A)\=\gamma^2\,\big(f^{abc}A^b_{\mu}(K^{-1})^{ad}f^{dec}
A^{e\mu}\ +\ D(N^2{-}1)\big)\ , \eeq
where $K^{-1}$ is the matrix inverse to the Faddeev-Popov operator
$K^{ab}$ in~(\ref{Landau}).
The so-called thermodynamic or Gribov parameter~$\gamma$ is
determined in a self-consistent way by the gap equation
\cite{Zwanziger1,Zwanziger2}
\begin{equation}
\frac{\partial \mathcal{E}_{vac}}{\partial \gamma}=0\ ,\label{gapeq}
\end{equation}
where $\mathcal{E}_{vac}$ is the vacuum energy given by
\begin{equation}
\exp\Big\{\frac{\im}{\hbar}\mathcal{E}_{vac}\Big\}\=\int\!D\Phi\
\exp\Big\{\frac{\im}{\hbar}S_{GZ}(\Phi)\Big\} \label{vacEn}
\end{equation}
pertaining to the Gribov-Zwanziger action
\cite{Sorella1,Sorella2,Sorella3}
\beq \label{GZact} S_{GZ}(\Phi)\=S_{FP}(\Phi)\ +\ M(A)\ . \eeq
Note that the functional $M(A)$ in~(\ref{FuncM}) is not invariant
under the BRST transformation (\ref{BRSTtr}) but trivially
satisfies the condition~(\ref{SoftBrC}) of soft BRST breaking
because of its independence on antifields.

The Gribov-Zwanziger action was intensively investigated in a series
of papers~\cite{Sorella1,Sorella2} where various quantum
properties of gauge models with this action have been studied. We
stress however that it was impossible in principle to establish the
gauge independence of physical quantities in these theories because
they were formulated practically in the Landau gauge~(\ref{Landau}) only, with except for covariant gauges in \cite{Sorella3}. Here,
we are going to clarify this crucial issue.

To this end, we discuss the Gribov-Zwanziger action (\ref{GZact})
for the one-parameter family of $R_\xi$ gauges,
\beq
\chi^a(A,B,\xi)
\=\partial^{\mu}A_{\mu}^a\ +\ \sfrac{\xi}{2}B^a
\label{gcYM} \eeq
with a real parameter $\xi$ interpolating between the Landau gauge ($\xi{=}0$)
and the Feynman gauge~($\xi{=}1$).
The Faddeev-Popov action is then written as
\beq
S_{FP}({\Phi,\xi})\=
S_0(A)\ +\ {\bar C}^a \partial^\mu D_\mu^{ab}C^b\ +\
(\partial^\mu A_\mu^a)B^a \ +\ \sfrac{\xi}{2} B^a B^a\ .\eeq
The Faddeev-Popov operator $K^{ab}$ is formally independent
of~$\xi$ if it is considered in acting of the space of the Yang--Mills fields $A_\mu^a$,
but the functional $M$ must be modified away from
$\xi{=}0$, already because $K^{ab}$ ceases to be
hermitian~\cite{Sorella1,Sorella2,Sorella3}.

To solve the problem of Gribov horizon definition here, we consider Hermitian augmented Faddeev-Popov operator for $R_\xi$ gauge,
\begin{equation}\label{FPaug}
\bar{K}^{ab}(\xi,A,B)=
\partial^{\mu}D_{\mu}^{ab}+ f^{acb}\frac{\xi}{2}B^c, \quad
\bigl(\bar{K}^{ab}(\xi)\bigr)^{+}=\bar{K}^{ab}(\xi),  \end{equation}
whose eigen-values
in the equation $\bar{K}^{ab}(\xi) u^b_n =
\lambda_n^au^a_n$  should be real and determine the Gribov region
$\Omega(\xi)$ as
\begin{eqnarray}
\nonumber \Omega(\xi)\equiv
\{A^a_{\mu},\;\partial^{\mu}A^a_{\mu}= - \frac{\xi}{2}B^a ,
K^{ab}>0\}
\end{eqnarray}
We \emph{suppose} that the proper eigen-values of the
$\bar{K}^{ab}(\xi)$ operator completely control  the ones of
non-Hermitian ${K}^{ab}(\xi)$ in such way, that the Gribov-Zwanziger
$R$-valued functional,
\begin{eqnarray}
\label{hermGZf}
M(A,B,\xi)=\gamma^2(\xi)\,\Big(f^{abc}A^{b{}T}_{\mu}(\bar{K}^{-1})^{ad}(A,B,\xi)f^{dec}
A^{e{}T{}\mu} + D(N^2{-}1)\Big)
\end{eqnarray}
 really determine Gribov region $\tilde{\Omega}(\xi)$. One should be noted, first, that thermodynamic Gribov parameter, $\gamma^2(\xi)$  should depend on gauge parameter, $\xi$, to be determined in self-consistent way from the Eqs. (\ref{gapeq}), (\ref{vacEn}) with $S_{GZ}(\Phi,\xi)$ and vacuum energy, $\mathcal{E}_{vac}(\xi)$. Second, we stress that
 so suggested introduction of   Gribov-Zwanziger
horizon functional is based on the representation of the Yang-Mills connection via transverse, $A^{a{}T}_{\mu}$, and longitudinal, $A^{a{}L}_{\mu}$, parts,
\beq\label{divide}
A^{a{}T}_{\mu} &= & \left(\delta_\mu^\nu - \frac{\partial_\mu\partial^\nu}{\partial^2}\right)A^{a{}}_{\nu} = A^{a{}}_{\mu}+ \xi\frac{\partial_\mu}{2\partial^2}B^{a{}},\\
A^{a{}L}_{\mu} &= & \frac{\partial_\mu\partial^\nu}{\partial^2}A^{a{}}_{\nu} = - \xi\frac{\partial_\mu}{2\partial^2}B^{a{}},
\eeq
introduced in \cite{Sorella3}, so that, $R_\xi$-gauge (\ref{gcYM}) is equivalent to the
pairs of conditions,
\beq\label{gauges}
\partial^\mu A^{a{}T}_{\mu} \ =\ 0, \qquad \partial^\mu A^{a{}L}_{\mu} = - \frac{\xi}{2}B^{a{}},
\eeq
As the consequence, the operator, $\bar{K}^{ab}(\xi,A,B)$, is nothing else, Faddeev-Popov  operator for the transverse components of Yang-Mills field, $A^{a{}T}_{\mu}$,
  \begin{equation}\label{FPatrans}
\bar{K}^{ab}(\xi,A,B) =
\partial^{\mu}(\partial_{\mu}^{ab} + f^{acb}A^{c{}T}_{\mu}) = {K}^{ab}(A^T).  \end{equation}

Now, we may state, following to \cite{Sorella3}, that the domain of integration with respect to the fields, $A^{a{}}_{\mu}$, in the path integral should be restricted to the region, $\tilde{\Omega}(\xi)$,
\beq\label{reginteg}
\tilde{\Omega}(\xi) = \big\{A^{a{}}_{\mu}| A^{a{}}_{\mu}= A^{a{}T}_{\mu} + A^{a{}L}_{\mu}; A^{a{}T}_{\mu}\in {\Omega}(\xi)\big\},
\eeq
 with $A^{a{}T}_{\mu}$ components of connections from only Gribov region,  ${\Omega}(\xi)$.

Of course, at least for small $\xi$ we suggest on smooth character of dependence of $\tilde{\Omega}(\xi)$ with Gribov region, ${\Omega}(0)$, for Landau gauge,
\beq\label{limreg}
\lim_{\xi \to 0}\tilde{\Omega}(\xi) = {\Omega}(0).
\eeq

 Now, we propose the
Gribov-Zwanziger action for Yang-Mills theories (\ref{clYM}) in the
$R_\xi$ gauge family~(\ref{gcYM}) as
\beq S_{GZ}(\Phi,\xi)\=S_{FP}(\Phi,\xi)\ +\ M(A,B,\xi)\ . \eeq
Because the BRST transformation (\ref{BRSTtr})
does not depend on the gauge fixing,
from~(\ref{hermGZf}) by continuity we can conclude that
\beq
\delta_B M(A,B,\xi)\neq 0 \qquad\longrightarrow\qquad
\delta_BS_{GZ}(\Phi,\xi)\neq 0\ .
\eeq
Let us recall our consistency condition~(\ref{BEqvM}), which takes the form
\beq \delta
M(A,B,\xi) \ \buildrel{!}\over{=}\
\sfrac12\frac{\delta M(A,B,\xi)}{\delta\Phi^A}\,
{\widehat F}_0^A\,\bar{C}^aB^a\,\delta\xi \ .\eeq
Since the right-hand side necessarily depends on the ghost, antighost
or auxiliary fields, it cannot match the left-hand side for our suggestion for $M$ in Eq. (\ref{hermGZf}).
Therefore, soft breaking of BRST symmetry is not consistent in $R_{\xi}$ gauges for Yang-Mills theory.

\section{Conclusions}\label{conclusion}

\noindent In the present paper we have considered a definition of
soft breaking of BRST symmetry in the field-antifield formalism using any
regularization scheme respecting gauge invariance. To this
purpose, we added a BRST `breaking functional'~$M$  to the gauge-fixed
action $S_{ext}$ which, in turn,  is constructed from an arbitrary
classical gauge-invariant action $\mathcal{S}_0$ by the BV method
rules.  The soft breaking of BRST symmetry was
determined by the analog of the quantum master equation,
$(M,M)=-2i\hbar \Delta M$. It was proved the non-invariance of the
integrand of
 vacuum functional under the  BRST transformations determined
by means of the functional, $(S_{ext}+\kappa M)$, for any value of
the real parameter $\kappa$.
 We have obtained all Ward identities for the generating
functional of Green's functions $Z$, of connected Green's
functions $W$ and of vertex functions $\Gamma$ both for dimensional-like regularization and for more general one, when $\delta(0) \ne 0$.
The Ward  identities were used to investigate the gauge dependence
of those functionals. It was argued that effective action  $\Gamma$ as well as the
S-matrix are on-shell gauge dependent. We were forced to claim
that a consistent quantization of gauge systems in the BV
formalism with the soft breaking of BRST symmetry does not exist.

We discussed  the Gribov-Zwanziger action
for the one-parameter family of $R_\xi$ gauges. To this aim we suggest the new form of the Gribov-Zwanziger horizon functional given in (\ref{hermGZf}) which is given on a base of Hermitian augmented Faddeev-Popov operator $\bar{K}^{ab}(\xi)$ in
(\ref{FPaug}) to be coinciding with the Faddeev-Popov operator constructed with respect only transverse component of Yang-Mills fields, suggested firstly in \cite{Sorella3}.  Already in this simple case,
the functional~$\Gamma$ turned out to depend on the gauge even on shell.
We are forced to conclude that a consistent quantization of gauge
theories with a soft breaking of BRST symmetry does not exist.

Our basic aim on this stage of finding a consistent quantum prescription to treat the theory with soft breaking of the BRST symmetry is to apply for   Yang--Mills and more general gauge theories with Gribov copies the BV formalism with composite fields. This perspective is now under our intensive consideration.

\section*{Acknowledgments}
\noindent
A.R. is grateful to V.Rubakov, S.Konstein and to the participants of the International Seminar QUARKS'2012 for valuable comments and discussion. The work is supported by  the LRSS grant 224.2012.2  as
well as by the RFBR grant 12-02-00121 and by the RFBR-Ukraine grant
11-02-90445.

\end{document}